\def\year{2018}\relax

\documentclass[letterpaper]{article} 
\usepackage{aaai18}  
\usepackage{times}  
\usepackage{helvet}  
\usepackage{courier}  
\usepackage{url}  
\usepackage{graphicx}  
\usepackage{subfigure}
\usepackage{booktabs}
\usepackage{paralist}
\usepackage{fix-cm}

\begin{document}

\title{AbuSniff: Automatic Detection and Defenses Against Abusive Facebook Friends}
\author{Sajedul Talukder\\
Florida Int'l University, Miami, USA\\
stalu001@fiu.edu
\And
Bogdan Carbunar\\
Florida Int'l University, Miami, USA\\
carbunar@gmail.com}
\maketitle

\begin{abstract}
Adversaries leverage social network friend relationships to collect sensitive data from users and target them with abuse that includes fake news, cyberbullying, malware, and propaganda. Case in point, 71 out of 80 user study participants had at least 1 Facebook friend with whom they never interact, either in Facebook or in real life, or whom they believe is likely to abuse their posted photos or status updates, or post offensive, false or malicious content. We introduce AbuSniff, a system that identifies Facebook friends perceived as strangers or abusive, and protects the user by unfriending, unfollowing, or restricting the access to information for such friends. We develop a questionnaire to detect perceived strangers and friend abuse. We introduce mutual Facebook activity features and show that they can train supervised learning algorithms to predict questionnaire responses.

We have evaluated AbuSniff through several user studies with a total of 263 participants from 25 countries. After answering the questionnaire, participants agreed to unfollow and restrict abusers in 91.6\% and 90.9\% of the cases respectively, and sandbox or unfriend non-abusive strangers in 92.45\% of the cases. Without answering the questionnaire, participants agreed to take the AbuSniff suggested action against friends predicted to be strangers or abusive, in 78.2\% of the cases.
AbuSniff increased the participant self-reported willingness to reject invitations from strangers and abusers, their awareness of friend abuse implications and their perceived protection from friend abuse.
\end{abstract}

\section{Introduction}

Social networks provide an ideal platform for abuse, that includes the collection and misuse of private user information~\cite{YS14,sextorsion,kontaxis2011detecting},
cyberbullying~\cite{SRHF17,kwan2013facebook}, and the distribution of offensive, misleading, false or malicious information~\cite{CBDL17,al2010taking,weimann2010terror,aro2016cyberspace}. The propensity of social networks towards such abuse has brought intense scrutiny and criticism from users, media, and politicians~\cite{MS08,CNBCabuse,NPRabuse,BBCSenate,BBCUK}.

Social networks like Facebook have made progress in raising user awareness to the dangers of making information public and the importance of deciding who can access it. However, many users still allow their Facebook friends to access their information, including timeline and news feed. This, coupled with the fact that people often have significantly more than 150 Facebook friends\footnote{Participants in our studies had up to 4,880 friends, $M$=305.} -- the maximum number of meaningful friend relationships that humans can manage~\cite{D92}) -- suggests that Facebook users may still be vulnerable to attacks.

To evaluate user perception of exposure to abusive friend behaviors, we designed 2 user studies (total $n$ = 80) where each participant had to evaluate 20 of their randomly selected Facebook friends. 65 of the 80 participants admitted to have at least 1 friend whom they perceived would abuse their status updates or pictures, and 60 of the participants had at least 1 friend whom they perceived would post abusive material (i.e., offensive, misleading, false or malicious). This is consistent with recent revelations of substantial abuse perpetrated through Facebook, including Cambridge Analytica's injection of content to change user perception~\cite{CAABC}, and Facebook's admission that in the past two years Russia-based operatives created 80,000 posts that have reached 126 million users in the US~\cite{BBCAbuse,BBCSenate}.

Further, 55 of the 80 participants admitted to have at least 1 Facebook friend with whom they have never interacted, either online or in person. Such {\it stranger} friends could be bots~\cite{VFDMF17} that passively collect sensitive user data and later use it against the user's best interest, as we also show through pilot study answers. This corroborates Facebook's recent estimate that 13\% (i.e., 270 million) accounts are either bots or clones~\cite{botsandclones}. Stranger friends can use the collected data to infer other sensitive user information~\cite{YS14}, identify ``deep-seated underlying fears, concerns'' by companies such as Cambridge Analytica~\cite{CAWired}, perform profile cloning~\cite{kontaxis2011detecting}, sextorsion~\cite{sextorsion}, identity theft~\cite{nosko2010all}, and spear phishing~\cite{GHWLCZ10} attacks.

These studies signal the need for defenses against abusive and stranger friends, that include restricting the abusers' access to user information, unfollowing them and even unfriending - removing them from the friend list. When asked directly, participants in our studies unfollowed and restricted access for abusers in 91.6\% and 90.9\% of the cases, respectively. When informed about the potential privacy risks posed by stranger friends, participants chose to unfriend or sandbox (block bi-directional communications with) such friends in 92.45\% of the cases.

\noindent
{\bf Contributions}.
We develop AbuSniff (Abuse from Social Network Friends), a system that evaluates, predicts and protects users against perceived friend abuse in Facebook. AbuSniff has the potential to mitigate the effects of abuse, and reduce its propagation through social networks and even its negative impact on social processes (e.g., electoral). We introduce the following contributions:

\begin{compactitem}

\item
Develop a friend abuse questionnaire that captures the user perception that a Facebook friend (1) is a stranger, (2) would publish abusive responses to pictures and status updates posted by the user, or (3) would publish and distribute offensive, misleading, false or potentially malicious information. Devise rules to convert identified abuse into defense actions.

\item
Propose the hypothesis that data recorded by Facebook can be used to predict the user perception of friend abuse. Introduce {\it mutual activity features} that quantify the Facebook recorded interactions between a user and her friend. Use supervised learning algorithms trained on these features to predict (1) user answers to the questionnaire, thus user perceived strangers and friend abuse and (2) the user willingness to take defensive actions against such friends.

\item
Implemented AbuSniff in Android (open source upon publication), and evaluated it through user studies with 263 participants from 25 countries and 6 continents.

\end{compactitem}

\noindent
{\bf Results}.
When using data we collected from 1,452 friend relationships ($n$=57), we found
that supervised learning algorithms trained on AbuSniff's mutual activity
features were able to predict the user answers to the questionnaire questions,
with an F-measure ranging between 69.2\% and 89.7\%.  Further, AbuSniff was
able to predict the cases where the users chose to ignore the suggested
defensive action against friends, with an F-Measure of 97.3\%.

In a user study ($n$ = 40) involving 1,200 Facebook friends, we found that without having to answer the questionnaire, participants accepted 78\% of AbuSniff's recommendations for defensive actions against abusive friends and strangers. In another study ($n$ = 31) AbuSniff increased participant self-reported willingness to reject invitations from perceived strangers and abusers, their awareness of friend abuse implications and perceived protection from friend abuse.

\section{Background and Model}
\label{sec:model}

We briefly summarize the relevant features of Facebook.  Facebook users form {\it friend} relationships. Each user has a {\it friend list} of other users with whom she has formed friend relationships.  The {\it timeline} (a.k.a wall, or profile) is Facebook's central feature, the place where the user can share her updates, photos, check-ins, and other activities (e.g., posting comments on a status or picture of a friend, confirming a new friend, etc).  These activities appear as {\it stories}, in reverse chronological order.  The timeline also includes friend activities that directly concern the user, e.g., their comments, status updates, notes or pictures that reference or include the user. This sensitive information is accessible by default by the user's friends. While users can control with whom they share each story, i.e., through the {\it audience selector} option, it is well known that they often use the default settings, see e.g.,~\cite{MJB12}. Further, a user's {\it news feed} shows stories created by her friends, groups, and subscribed events. Stories are sorted based on various features, e.g., post time and type, poster popularity.

\subsection{Adversary Model}
\label{sec:model:adversary}

We consider adversaries who leverage the following mechanisms to perpetrate abuse through Facebook:

\begin{compactitem}

\item
{\bf Privacy abuse}.
Collect sensitive information (profiles, photos, friend lists, locations
visited, opinions) posted by friends on their timelines or take screenshots of stories. The adversary can then use this data to infer more sensitive information~\cite{YS14}, initiate sextorsion~\cite{sextorsion}, perform profile cloning~\cite{kontaxis2011detecting}, identity theft~\cite{nosko2010all}, and spear phishing~\cite{GHWLCZ10} attacks. Facebook estimates that 13\% (i.e., 270 million) of their accounts are either bots or clones~\cite{botsandclones}.

\item
{\bf Timeline abuse}.
Post abusive replies to stories (e.g., status updates, photos) posted by friends on their timeline. The abusive replies appear on the timeline of the victim, where the original stories were posted.

\item
{\bf News-feed abuse}.
The adversary posts abusive material on his timeline, which is then propagated to the news feed of his friends. Abusive information includes material perceived to be offensive, misleading, false, or malicious. Facebook revealed that Russia-based operatives created 80,000 posts that have reached 126 million US users~\cite{BBCAbuse,BBCSenate}.

\end{compactitem}

\subsection{Restrictive Actions Against Friends}
\label{sec:model:actions}

AbuSniff leverages several defense mechanisms provided by Facebook to protect the user against strangers and abusive friends: {\bf unfollow} -- stories subsequently posted by the friend in his timeline no longer appear in the user's news feed, {\bf restrict} -- stories published by the user in her timeline no longer appear in the friend's news feed, and {\bf unfriend} -- remove the friend from the user's list of friends.

Further, we introduce the {\bf sandbox} defense option, a combination of unfollow and restrict: the user and her friend no longer receive stories published by the other. Unlike unfriending, sandboxing will not remove the user and her friend from each other's friend lists.

\section{Research Objectives}
\label{sec:objectives}

We study several key questions about friend based abuse:

\begin{compactitem}

\item
{\bf (RQ1)}: Are perceived strangers and friend abuse real problems in Facebook?  

\item
{\bf (RQ2)}: Are Facebook users willing to take defensive actions against abusive friends?

\item
{\bf (RQ3)}: Does AbuSniff have an impact on the willingness of users to take defensive actions on Facebook friends, and is this willingness impacted by the type of abuse perpetrated by the friend and the suggested defensive action?

\item
{\bf (RQ4)}: Can AbuSniff predict abusive friends and the defenses that users are willing to take against such friends?

\item
{\bf (RQ5)}: Does AbuSniff impact user awareness of stranger and abusive friends, and their perception of safety from such friends?

\end{compactitem}

\noindent
In order for AbuSniff to be relevant, RQ1, RQ2 and RQ5 need to have positive answers. The answer to RQ3 will impact the design of AbuSniff, while a positive answer to RQ4 will indicate that systems can be built to automatically defend users against friend abuse.

\section{The AbuSniff System}
\label{sec:abusnf}

\begin{figure}
\centering
\includegraphics[width=0.47\textwidth]{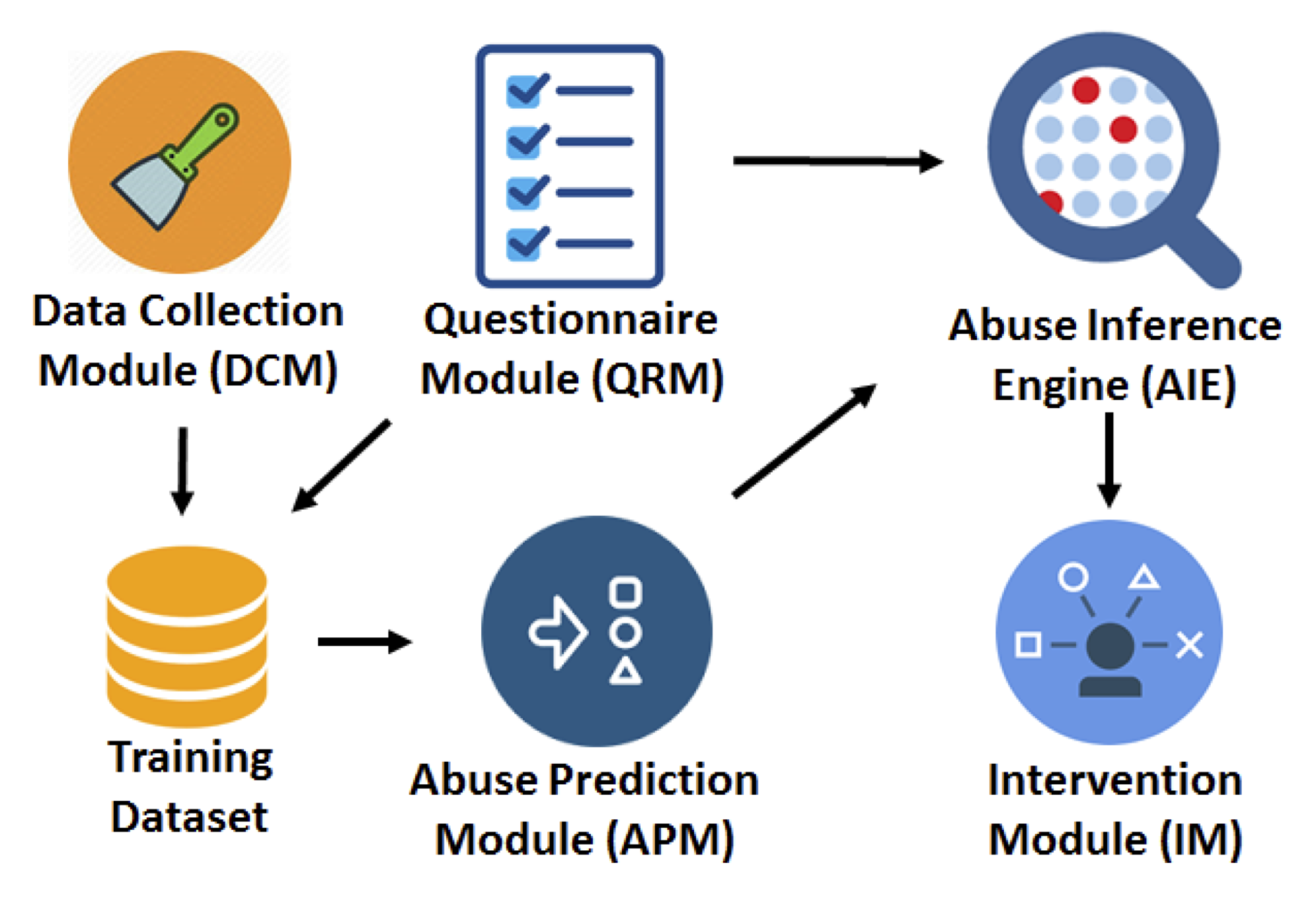}
\caption{AbuSniff system architecture. The QRM module delivers the questionnaire.  The DCM module collects the user responses and also Facebook data concerning the relationship with each friend. The APM module uses the collected data to predict the responses to the questionnaire. The AIE module uses the output of the QRM or APM to identify abusive friends, and the IM module asks the user to take a protective action against them.}
\label{fig:abusniff}
\end{figure}

We have designed the AbuSniff system to help us investigate these questions. AbuSniff is a mobile app that asks the user to login to her Facebook account. As illustrated in Figure~\ref{fig:abusniff}, AbuSniff consists of modules to collect user responses and data, predict user responses, identify abusive friends, and recommend defensive actions. In the following, we describe each module.

\subsection{The Questionnaire Module (QRM)}
\label{sec:abusnf:qrm}

We have designed a questionnaire intended to capture the user perception of (potentially) abusive behaviors from friends in Facebook. Since Facebook users tend to have hundreds and even thousands of friends, we decided to present the questionnaire for each of only a randomly selected subset of the user's friends.  One design goal was that the questions should help identify the perceived use of the abusive mechanisms listed in the adversary model. To ensure a simple navigation of the questionnaire, we further sought to fit all the questions on a single screen for a variety of popular smartphones. We have designed the questionnaire through an iterative process that included a focus group and a pilot study with 2 K-8 teachers, 1 psychologist, 8 students, 1 dentist and 1 homemaker (8 female and 5 male).

\begin{figure*}
\centering
\includegraphics[width=0.99\textwidth]{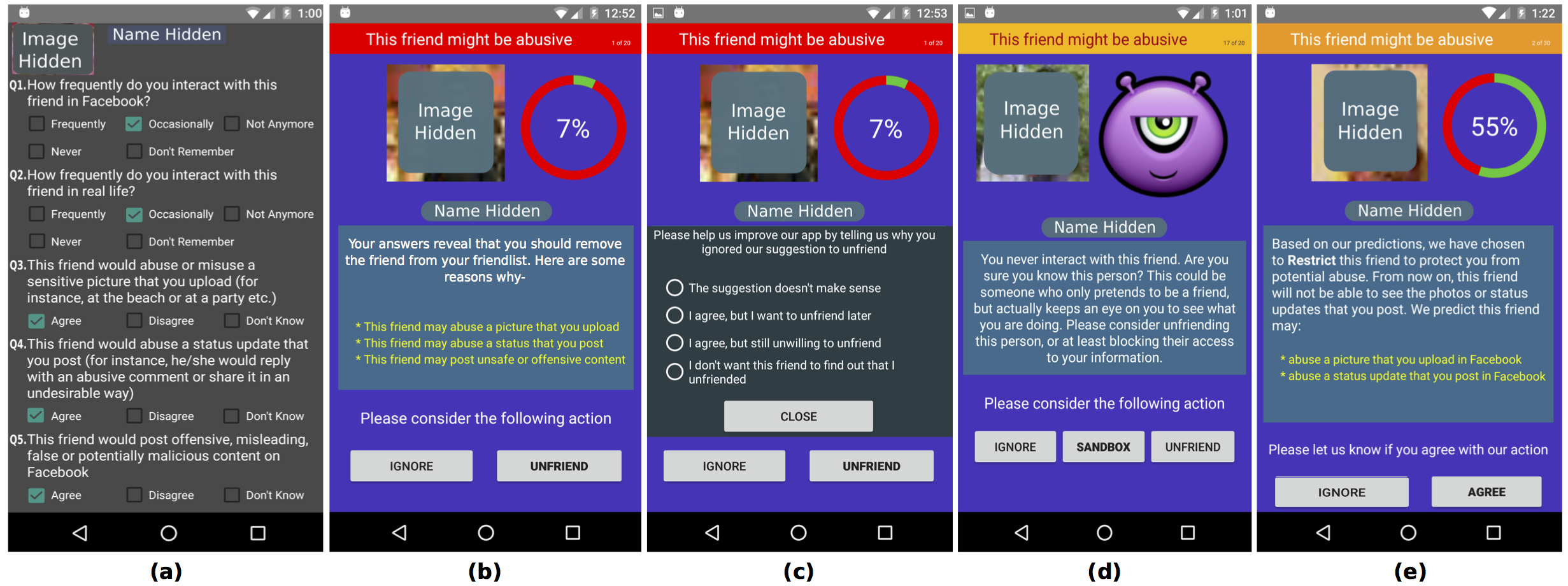}
\caption{Anonymized screenshots of the Android AbuSniff app:
(a) QRM questionnaire. The first two questions identify stranger friends, questions 3 and 4 identify perceived timeline abuse and question 5 identifies perceived news feed abuse.
(b) The IM UI asking the user to unfriend an abusive friend also explains the reasons for the action, according to the questionnaire responses.
(c) The IM UI asking the user to explain the reasons for the unwillingness to unfriend in the previous screen.
(d) The ``unfriend or sandbox'' UI for privacy abuse: sandboxing isolates but does not unfriend or notify the friend.
(e) The UI of the autonomous AbuSniff asking user confirmation to restrict the access of a friend predicted to be a timeline abuser.}
\label{fig:abusniff:screenshot}
\end{figure*}

Figure~\ref{fig:abusniff:screenshot}(a) shows a snapshot of the resulting questionnaire, that consists of 5 questions.  The first two questions (Q1) ({\it How frequently do you interact with this friend in Facebook}) and (Q2) ({\it How frequently do you interact with this friend in real life}) determine the user's frequency of interaction with the friend, on Facebook and in real life. The options are ``Frequently'', ``Occasionally'', ``Not Anymore'' (capturing the case of estranged friends), ``Never'' and ``Don't Remember''. We are particularly interested in the ``Never'' responses.

After answering ``Never'' for Q1 for a friend, participants in the focus group explained that they have never initiated conversations with the friend and are either not aware of or interested in communications initiated by the friend, e.g.,

{\it ``I never did chat with him, he never commented on my photos or any shared thing. He never puts a like [sic].''}

{\it ``I never like or comment on his post, I never chat with him. [..] Actually I do not notice if he likes my posts. But I do not do [sic] any interaction.''}

For question Q2, participants agreed that they have never met in real life friends for whom they answered ``Never''. Reasons for accepting the friend invitations from such friends include {\it ``he is a friend of my friend and my friend met him in real life''}, and {\it ``she is from my same [sic] college''}. This suggests that friends with whom the user has never interacted in Facebook and in real life, may be {\it strangers}. Such strangers may exploit Facebook affordances (e.g., claim college education) to befriend victims.

The next two questions identify perceived timeline abusers, i.e., (Q3) {\it This friend would abuse or misuse a sensitive picture that you upload} and (Q4) {\it This friend would abuse a status updated that you upload}. The possible responses are ``Agree'', ``Disagree'' and ``Don't Know''. After answering ``Agree'' for Q3, participants shared several stories of abuse, e.g.,

{\it ``Once this friend has downloaded my photo and then opened a fake Facebook account, like with that picture.''}, and

{\it ``This friend has posted a bad comment in one of my photos. That was my wedding photo. I felt so offended.''}

Participants who answered ``Agree'' for a friend on Q4 shared other stories of abuse, e.g.:

{\it ``This friend posted a bad comment on my post and from that post there was other bad stuff posted on my wall.''}, and

{\it ``Once I posted a sad status update because I was feeling frustrated. But this friend then posted a trolling comment on my post.''}

The last question (Q5) {\it This friend would post offensive, misleading, false or potentially malicious content on Facebook} identifies perceived news-feed abusers. Stories shared by participants who answered ``Agree'' on Q5 include:

{\it ``This friend bothered friends by bad posts [..] The posts were against my own ideas [sic].''}, and

{\it ``I have often seen this friend sharing fake news. Sometimes she posts so much bogus stuff that my news feed gets flooded.''}

These examples show that privacy and security abuses occur in the real life interactions of Facebook users and their friends. The following AbuSniff modules seek to predict the user perception of abuse and convert it into defensive actions that users will consider appropriate.

\begin{table}[t!]
\centering
\resizebox{0.47\textwidth }{!}{%
\begin{tabular}{l l l l l l l }\hline
& \textbf{Q1} & \textbf{Q2} & \textbf{Q3} & \textbf{Q4} & \textbf{Q5} & \textbf{Action}\\ \hline
1 & Never & Never & !Agree & !Agree & !Agree & Unfriend/\\
  &       &       &              &              &              & Sandbox\\
2 & Never & Never & $\ast$ & $\ast$ & $\ast$ & Unfriend\\
\midrule
3 & Never & !Never & Agree & Agree & Agree & Unfriend\\
4 & !Never & Never & Agree & Agree & Agree & Unfriend\\
5 & Never & !Never & Agree & !Agree & Agree & Unfriend\\
6 & Never & !Never & !Agree & Agree & Agree & Unfriend\\
7 & !Never & Never & Agree & !Agree & Agree & Unfriend\\
8 & !Never & Never & !Agree & Agree & Agree & Unfriend\\
9 & !Never & !Never & Agree & Agree & Agree & Unfriend\\
10 & !Never & !Never & Agree & !Agree & Agree & Unfriend\\
11 & !Never & !Never & !Agree & Agree & Agree & Unfriend\\
\midrule
12 & !Never & !Never & Agree & Agree & !Agree & Restrict\\
13 & !Never & !Never & Agree & !Agree & !Agree & Restrict\\
14 & !Never & !Never & !Agree & Agree & !Agree & Restrict\\
\midrule
15 & !Never & !Never & !Agree & !Agree & Agree & Unfollow\\
\midrule
16 & $\ast$ & $\ast$ & $\ast$ & $\ast$ & $\ast$ & NOP\\
\bottomrule
\end{tabular}
}\caption{Set of rules to convert questionnaire responses to defensive actions. Like firewall filters, the first matching rule applies. !$A$ denotes any response different from $A$. NOP $=$ no operation.}
\label{table:rules}
\end{table}

\subsection{The Abuse Prediction and Data Collection Modules }
\label{sec:abusnf:apm}

We investigate the ability of a supervised learning approach to provide an affirmative answer to question RQ4 ({\it can AbuSniff predict the abusive friends and the defenses that users are willing to take against such friends?}).

We introduce 7 {\it mutual activity} features, based on the Facebook data shared by a user $U$ and a friend $F$: (1) {\bf mutual post count}: the number of stories posted by either $U$ or $F$, on which the other has posted a comment, (2) {\bf common photo count}: the number of photos in which both $U$ and $F$ are tagged together, (3) {\bf mutual friend count}: the number of common friends of $U$ and $F$, (4,5) {\bf same current city and hometown}: boolean values that are true when $U$ and $F$ live in the same city and are from the same place, (6,7) {\bf common study and work count}: the total number of places where $U$ and $F$ have studied and are employed together, respectively.

The abuse prediction module (APM) uses supervised learning algorithms trained on these features, and previously collected questionnaire responses and user decisions, to predict the user's answers to the QRM questionnaire and the user's reactions to suggested actions. We report accuracy results in the evaluation section.

The Data Collection Module (DCM) collects Facebook data from the user and her evaluation friends, as well as user provided input (e.g., responses from the QRM, choices from the IM) and timing information. AbuSniff uses this data to make local decisions and partially reports it to our server for evaluation purposes.

\subsection{The Abuse Inference Engine (AIE)}
\label{sec:abusnf:aie}

\cite{VK14} found that to mitigate risks, prudent social network users (i.e., graduate students) used a variety of risk management techniques that include limiting the recipients of posts, hiding friends from their news feed, and unfriending friends. AbuSniff seeks to provide similarly safe social interactions to regular social network users. To this end, the abuse inference engine (AIE) takes as input the responses collected by the QRM or predicted by the APM, and outputs suggested actions from the set $\{$ ``unfriend'', ``unfollow'', ``restrict access'', ``sandbox'', ``ignore''$\}$. 

AIE uses the rules shown in Table~\ref{table:rules}, applied on a first match basis: rule $r$ is evaluated only if all the rules 1 to $r$ - 1 have failed. The first 15 rules detect restrictive actions; if none matches, the last rule decides that the friend is not abusive (i.e., ignore). Initially, we took a hard stance against abuse: a friend who scores negatively on any 2 out of the 5 questions (i.e., assigned ``Never'' in any of the 2 first questions, ``Agree'' in any of the last 3 questions) should be unfriended (rules 1-11). However, AIE outputs less restrictive actions against friends with whom the user has interacted both in Facebook and in real life, and is either only a timeline abuser (restrict, rules 12-14) or only a news-feed abuser (unfollow, rule 15). We evaluate and adjust these rules in the evaluation section.

\subsection{The Intervention Module (IM)}
\label{sec:abusnf:im}

To help us answer the key research questions RQ2 and RQ3, we have designed a
user interface that asks the user to take a defensive action against each
friend detected as abusive by the AIE module. The action, i.e., unfriend,
restrict, unfollow, is determined according to the rule matched in
Table~\ref{table:rules}.

Figure~\ref{fig:abusniff:screenshot}(b) shows a snapshot of the ``unfriend''
recommendation. The UI further educates the user on the meaning of the action,
and lists the reasons for the suggestion, based on the questionnaire responses
that have matched the rule, see Figure~\ref{fig:abusniff:screenshot}(a).

The user is offered the option to accept or ignore the suggestion.  If the user
chooses to ignore the suggestion, the IM module asks the user (through a
PopupWindow) to provide a reason, see Figure~\ref{fig:abusniff:screenshot}(c).
We have conducted a focus group with 20 participants in order to identify
possible reasons for ignoring ``unfriend'' recommendations. They include ``the
suggestion does not make sense'', ``I agree, but I want to unfriend later'',
``I agree but I am still unwilling to unfriend'', and ``I don't want this
friend to find out that I unfriended'', see
Figure~\ref{fig:abusniff:screenshot}(c). We did not include an open text box,
as we did not expect that participants will type an answer on a mobile device.

The IM module educates users about the meaning and dangers of having a stranger as a friend, see Figure~\ref{fig:abusniff:screenshot}(d). It also offers the option to ``sandbox'' such friends. According to the rules of Table~\ref{table:rules}, IM also suggests unfollowing or restricting friends who are abusive in only one direction of their communications. Figure~\ref{fig:abusniff:screenshot}(e) shows a snapshot of the restrict screen, its meaning and reasons for selection.

\section{User Study}
\label{sec:study}

\begin{figure}[t]
\centering
\includegraphics[width=0.45\textwidth]{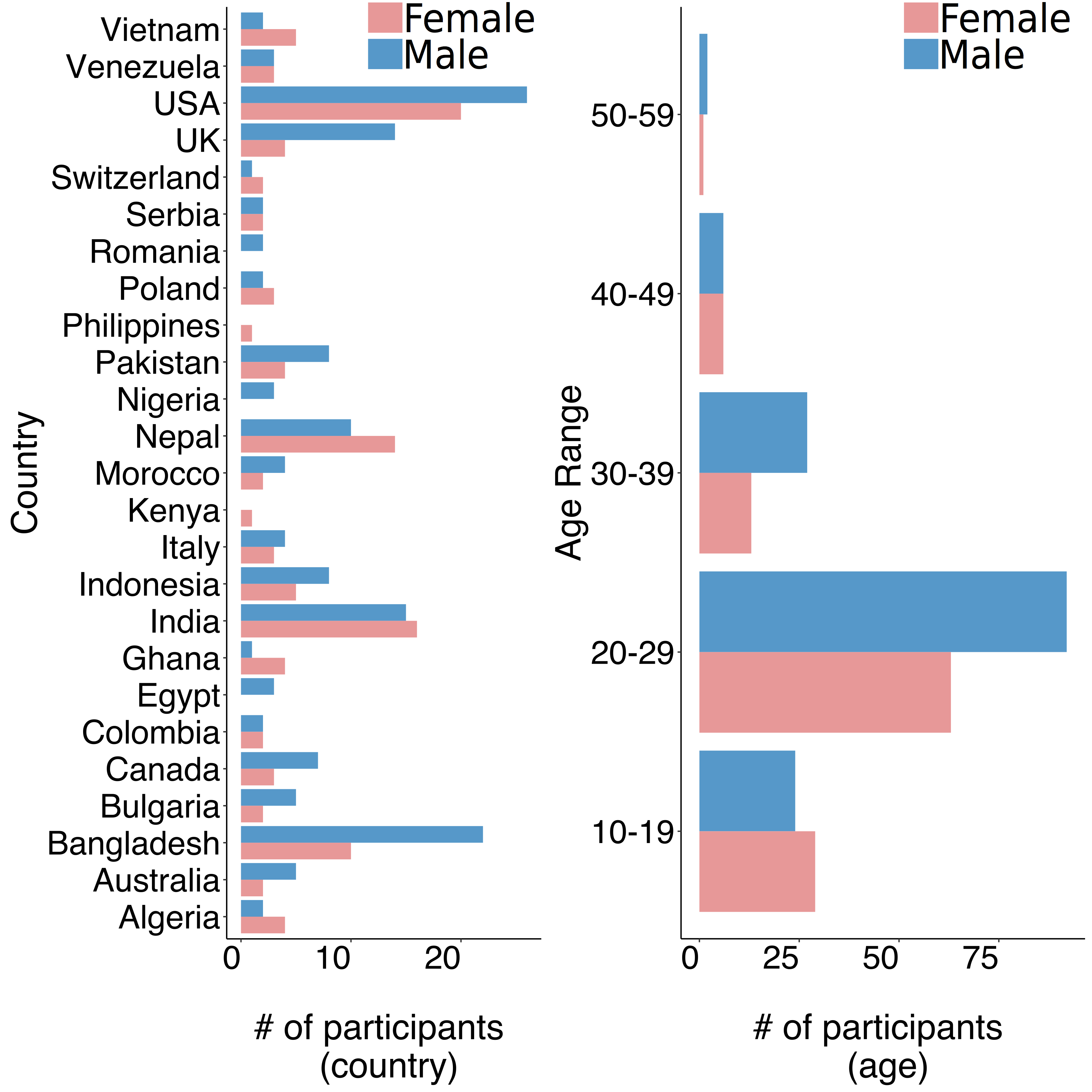}
\caption{Participant demographics.
(country) Distribution of the 25 countries of residence by gender.
(age) Distribution of age range by gender.
}
\label{fig:user:statistics}
\end{figure}

We have conducted several user studies to answer our key research questions. In the following we describe the participant recruitment procedure, the experiment design, and techniques we used to ensure data quality.

We have recruited 325 participants from JobBoy~\cite{JobBoy}, during 7 studies conducted between August 2016 and October 2017. The jobs we posted asked the participants to install AbuSniff from the Google Play store, use it to login to their Facebook accounts and follow the instructions on the screen. A participant who successfully completes the app, receives on the last screen a code required for payment. We have paid each participant \$3, with a median job completion time of 928s (SD = 420s).

We have only recruited participants who have at least 30 Facebook friends, had access to an Android device, and were at least 18 years old. Further, we have used the following mechanisms to ensure the quality of the data collected.

\noindent
$\bullet$
{\bf Attention-check screen}.
To ensure that the participants pay attention and are able to understand and follow simple instructions in English, AbuSniff includes a standard attention-check screen at the beginning of the app.

\noindent
$\bullet$
{\bf Bogus friends}.
To detect participants who answer questions at random, we used ``bogus friends'': 3 fake identities (2 female, 1 male) that we included at random positions in the AbuSniff questionnaire. We have discarded the data from participants who answered Q1 and Q2 for the bogus friends, in any other way than ``Never'' or ``Don't Remember''.

\noindent
$\bullet$
{\bf Timing information}.
We have measured the time taken by participants to answer each questionnaire question and to make a decision on whether to accept or ignore the suggested action. We have discarded data from participants whose average response time was below 3s.

We have used these mechanisms to discard 62 of the recruited 325 participants. The following results are shown over the remaining 263 participants. Figure~\ref{fig:user:statistics} shows the distribution of the country of origin (left) and age (right), by gender, over these participants. The 151 male and 112 female participants are from 25 countries (top 5: US, Bangladesh, India, Nepal and UK) and 6 continents, and are between 18-52 years old (M = 23, SD = 7.22).

\subsection{Ethical Data Collection}
\label{sec:study:ethics}

We have developed our protocols to interact with participants and collect data in an ethical, IRB-approved manner (Approval \#: IRB-16-0329-CR01). The 54 participants from whose friends we collected mutual activity features, were made aware and approved of this data collection step. We have collected minimalistic Facebook data about only their investigated friend relationships.  Specifically, we have only collected the counts of common friends, posted items, studies and workplaces, and boolean values for the same current city and hometown, but not the values of these fields. Further, we have only collected anonymized data, and the automated AbuSniff version {\it never} sends this data from the user's mobile device. AbuSniff only uses the data to make two predictions (the type of abuse and whether the user will take the suggested action, then erases the collected Facebook data.

\section{Results}
\label{sec:evaluation}

\subsection{Abuse Perception and Willingness to Defend}
\label{sec:evaluation:abuse}

\begin{figure}
\centering
\includegraphics[width=0.47\textwidth]{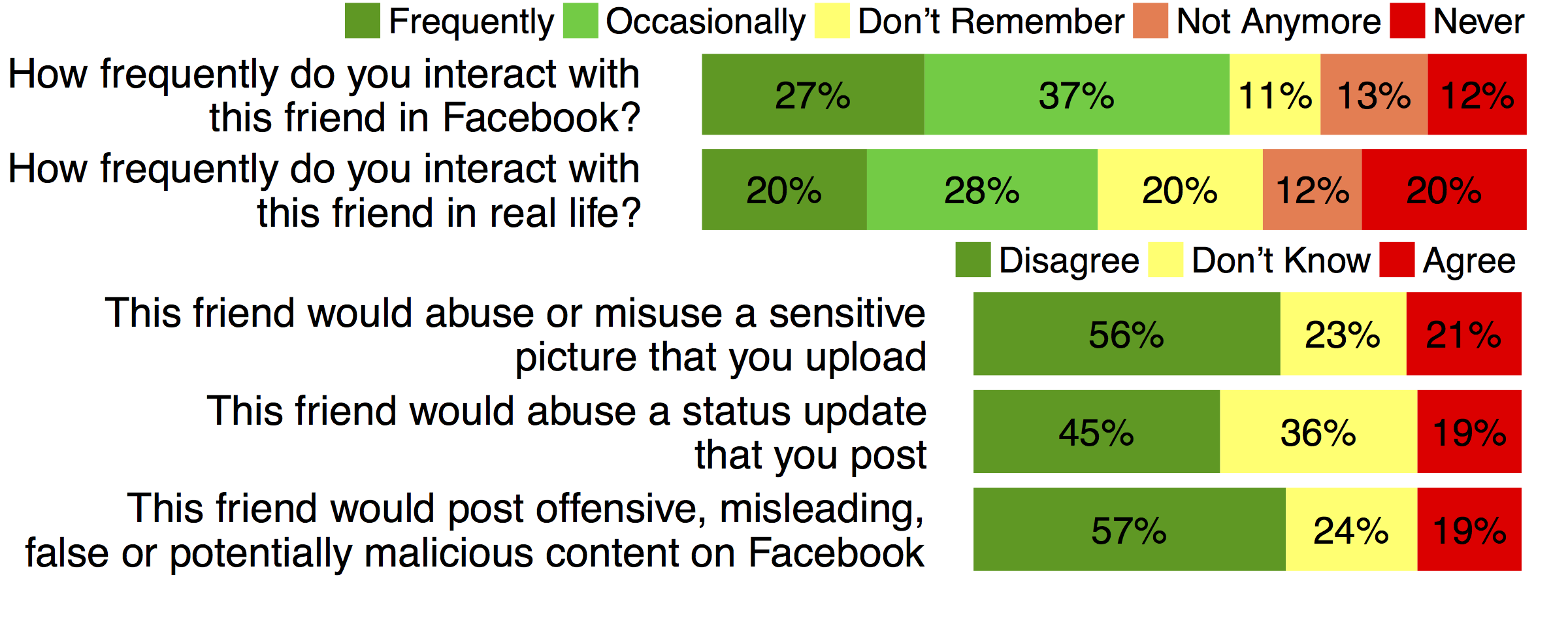}
\caption{Distribution of responses for the friend abuse questionnaire over 1,600 Facebook friend relationships. The red sections correspond to potential strangers or abusive friends.}
\label{fig:questionnaire:stats}
\vspace{-15px}
\end{figure}

We developed 2 preliminary studies ($n$ = 20 and $n$ = 60) to evaluate the extent of the user perception of stranger friends and friend abuse in Facebook (RQ1) and the willingness of users to accept defensive actions against friends considered to be abusive (RQ2 and RQ3). To this end, AbuSniff used the QRM, DCM, AIE and IM modules, see Figure~\ref{fig:abusniff}. Further, AbuSniff randomly selected 20 Facebook friends of each participant, asked the participant to answer the questionnaire for each friend, then asked the participant to take a defensive action against the friends detected to be abusive, or provide a reason for ignoring the suggested action.

Figure~\ref{fig:questionnaire:stats} shows the distribution of the responses for each of the 5 questions from the 1,600 friend relationships (20 from each participant). Further, 64 of the 80 participants stated that they have at least one friend with whom they have never interacted in Facebook, while 73 of the participants had at least one friend with whom they have never interacted in real life. 68 of the participants had at least 1 friend whom they perceived would abuse their photos, 62 of the participants have at least 1 friend who would abuse their status updates, and 62 have at least 1 friend who would post abusive content.

\noindent
{\bf Gender and age impact}.
In terms of having at least 1 friend perceived as abusive, Chi-square tests revealed no significant difference between genders on any of the 5 questions. Similarly, Chi-square tests revealed no significant differences between the age groups of under 30 years old and above 30 years old participants (61 vs 19 participants), on questions 1, 2 and 4.  However, participants under 30 are significantly more likely ($\chi^2$ = 4.417, df = 1, p = 0.03) to have at least 1 friend whom they perceive would abuse a photo they post, than participants over 30 (52 out of 61 vs 12 out of 19). Younger participants were also more likely to answer that they have at least 1 friend who would post offensive, misleading, false or potentially malicious content (50 out of 61  vs 10 out of 19, $\chi^2$ = 6.64, df = 1, p = 0.01).

\begin{figure}
\centering
\includegraphics[width=0.45\textwidth]{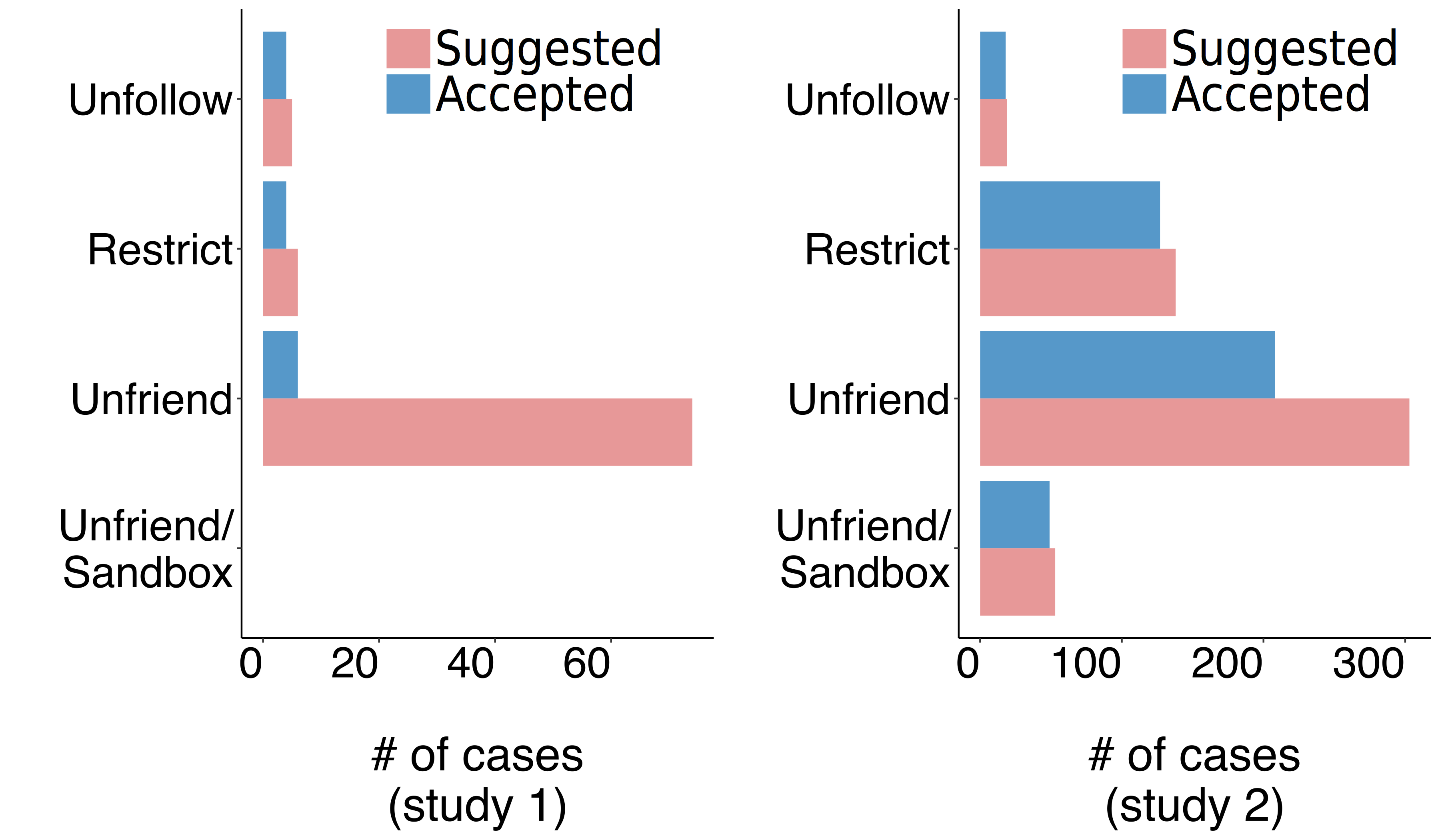}
\caption{
Recommendation vs. acceptance in study 1 ($n$ = 20) and study 2 ($n$ = 60). {\bf The ``sandbox'' option and user education were effective}: 92\% of the suggested ``unfriend or sandbox'' suggestions were approved by participants.}
\label{fig:acceptance}
\vspace{-15px}
\end{figure}

\noindent
{\bf Willingness to Defend Against Abuse}.
In the first of the above 2 studies ($n$ = 20, 400 investigated friend relationships), AbuSniff identified 85 abusive friend relations. Of these, AbuSniff recommended 74 to unfriend, 6 to restrict and 5 to unfollow.
The results are summarized in Figure~\ref{fig:acceptance}(a). 4 out of the 6 recommended restrict friends were restricted, and 4 out of the recommended 5 were unfollowed. However, only 6 out of 74 recommended unfriend were unfriended. In 55 of the 68 (74 - 6) unfriended friends, the participants believed that our warning was correct. However, they refused to unfriend because either they were not ready to unfriend them at that time (18 of the 55 cases), they still wanted to keep those abusive friends (11 cases), or they were afraid that this action will be observable by the abuser (26 cases).

\noindent
{\bf The ``Sandbox'' Effect}.
To address the fear of being observed by the unfriended friend, we have relaxed rule 1 in Table~\ref{table:rules}, to give the user the option to either sandbox or unfriend a non-abusive stranger. A sandboxed friend can no longer harm the user, as all Facebook communication lines are interrupted. Sandboxing achieves this without severing the friend link, thus is not observable by the friend. Further, we have modified AbuSniff's UI to educate the user through a description of the harm that strangers can perform, and of the defenses that the user can take against such friends. Figure~\ref{fig:abusniff:screenshot}(d)) shows a snapshot of the modified UI screen that offers the sandbox alternative to unfriending strangers.

The second user study described above ($n$ = 60) evaluated the updated AbuSniff, that identified a total of 513 abusive friend relations. Figure~\ref{fig:acceptance}(study 2) shows that AbuSniff recommended 303 to unfriend, 53 to unfriend or sandbox, 138 to restrict and 19 to unfollow.  Consistent with the first study, 18 of the 19 unfollow and 127 of the 138 restrict suggestions were accepted. In contrast to the first study, 49 of the 53 ``unfriend or sandbox'' suggestions were accepted. In addition, 208 of 303 ``pure'' unfriend recommendations were accepted, again a significant improvement over the first study (6 out of 74). Only 5 out of the 95 ignored unfriend recommendations were due to the participant not believing our recommendation.

\subsection{Efficacy of Abuse Prediction}
\label{sec:evaluation:prediction}

\begin{table}[t]
\centering
\resizebox{0.47\textwidth}{!}{%
\small
\textsf{
\begin{tabular}{c l l l l l l}
\toprule
\textbf{Question} &\textbf{Precision} & \textbf{Recall} & \textbf{F-Measure} & \textbf{Class} \\
\midrule
    & 0.983  & 1.000  & 0.992  & Frequently \\
    & 0.928 & 0.897 & 0.912 & Occasionally \\
Q1  & 0.962  & 0.797 & 0.872 & Not Anymore \\
(RF) & \textbf{0.818} & \textbf{0.920} & \textbf{0.866} & \textbf{Never} \\
    & 0.934 & 0.898 & 0.916 & Don't Remember \\
    & 0.917 & 0.914 & 0.914 & Weighted Avg. \\
\midrule
    & 0.966 & 0.905  & 0.934  & Frequently \\
    & 0.893 & 0.869 & 0.881 & Occasionally \\
Q2  & 0.893 & 0.877 & 0.885 & Not Anymore \\
(RF) & \textbf{0.865} & \textbf{0.932} & \textbf{0.897} & \textbf{Never}\\
    & 0.907 & 0.911 & 0.909 & Don't Remember \\
    & 0.902 & 0.900 & 0.900 & Weighted Avg. \\
\midrule
    & \textbf{0.725}  & \textbf{0.792} & \textbf{0.757}  & \textbf{Agree} \\
Q3  & 0.820  & 0.793 & 0.806 & Disagree \\
(DT) & 0.810 & 0.791 & 0.800 & Don't Know \\
    & 0.794 & 0.792 & 0.793 & Avg. \\ \hline
    & \textbf{0.662} & \textbf{0.725} & \textbf{0.692}  & \textbf{Agree} \\
Q4  & 0.791  & 0.778 & 0.785 & Disagree \\
(DT) & 0.857  & 0.844 & 0.851 & Don't Know \\
    & 0.805 & 0.803 & 0.804 & Avg. \\
\midrule
    & \textbf{0.794} & \textbf{0.765} & \textbf{0.780}  & \textbf{Agree} \\
Q5  & 0.837 & 0.845 & 0.841 & Disagree \\
(RF) & 0.830  & 0.842 & 0.836 & Don't Know \\
    & 0.824 & 0.824 & 0.824 & Avg. \\
\bottomrule
\end{tabular}
}}
\caption{Precision, recall and F-measure of APM for questions Q1 (RF), Q2 (RF), Q3 (DT), Q4 (DT) and Q5 (RF).
}
\label{table:prediction}
\end{table}

To answer key question RQ4, in a fourth study we used AbuSniff to collect a subset of Facebook data from 1,452 friend relationships of 54 participants. We have computed the 7 mutual activity features of the 54 participants and the 1,452 friends, and used 10-fold cross validation to evaluate the ability of the abuse prediction module (APM) to predict questionnaire responses and user defense decisions.

As shown in Figure~\ref{fig:questionnaire:stats}, the distribution of the answers to the 5 questions of the questionnaire was not balanced.  To address this imbalance, we have duplicated tuples from the minority classes up to the the number of the majority class. We have ensured that duplicates appear in the same fold, to prevent testing on trained tuples. We have used Weka 3.8.1~\cite{Weka} to test several supervised learning algorithms, including Random Forest (RF), Decision Trees (DT), SVM, PART, MultiClassClassifier, SimpleLogistic, K-Nearest Neighbors (KNN) and Naive Bayes, but report only the best performing algorithm.

\noindent
{\bf Predicting questionnaire answers}.
Table~\ref{table:prediction} shows the precision, recall and F-measure achieved by the best performing supervised learning algorithm for each of the questionnaire questions (Q1-Q5). The RF classifier achieved the best F-measure for questions Q1, Q2 and Q5, while the DT classifier achieved the best F-measure for Q3 and Q4. We observe a higher F-measure in predicting answers to the questions that suggest stranger friends (Q1 and Q2) than in predicting answers to the questions that suggest abuse (Q3-Q5). This is not surprising, as the mutual activity features are more likely to predict online and real life closeness.

\noindent
{\bf Predicting the user decision}.
We have evaluated the ability of APM to predict the defense action that the user agrees to implement, according to the 5 possible classes: ``unfriend'', ``restrict'', ``unfollow'', ``sandbox'', and ``ignore''. APM achieved the best performance with the RF classifier. Table~\ref{table:prediction:action:matrix} shows the confusion matrix for APM with RF, over the 10-fold cross validation performed on the 1,452 friend instances. While the overall F-Measure is 73.2\%, APM achieved an F-measure of 97.3\% when predicting the ``ignore'' option.

\begin{table}
\centering
\resizebox{0.47\textwidth}{!}{%
\textsf{
\begin{tabular}{l l l l l |l}
\toprule
\multicolumn{5}{c|}{\textbf{Classified As}}\\
\textbf{Unfriend} & \textbf{Sandbox} & \textbf{Restrict} & \textbf{Unfollow} & \textbf{Ignore} & \textbf{Decision} \\
\midrule
\textbf{882}     & 13     & 10     & 13     & 3   & \textbf{Unfriend} \\
103     & 27     & 1     & 1     & 3    & Sandbox \\
77     & 1     & 6     & 0     & 1    & Restrict \\
79     & 3     & 0     & 6     & 0    & Unfollow \\
5     & 0     & 0     & 0   & \textbf{218}     & \textbf{Ignore} \\
\bottomrule
\end{tabular}%
}}
\caption{APM confusion matrix for predicting user decisions. The rows show participant decisions, the columns show APM predictions during the experiment. {\bf AbuSniff will leverage APM's high precision (96.9\%) and recall (97.8\%) for the ``ignore'' action, to decide which abusive friends to ignore.}}
\label{table:prediction:action:matrix}
\end{table}

\noindent
{\bf Feature rank}.
The most informative features in terms of information gain were consistently among the mutual post count, mutual friend count and mutual photo count; same hometown and common study count were the least informative features. We found correlations between the common photo count and mutual post count (Pearson correlation coefficient of 0.65), mutual friend count and mutual photo count (Pearson correlation coefficient of 0.57), and mutual post count and mutual friend count (Pearson correlation coefficient of 0.45). The rest of the features had insignificant positive or negative correlations.

\subsection{AbuSniff in the Wild}
\label{sec:evaluation:wild}

To evaluate the autonomous AbuSniff live, on real users, we have replaced the questionnaire delivery module (QRM) with the abuse prediction module (APM). AbuSniff then asks the user to either accept or ignore the APM predicted defense action, only for the friends for whom the APM predicts that the user will defend against.

We have recruited 49 participants to evaluate their reaction to the predictions of the APM module. We have discarded 9 participants who failed the data quality verification tests previously described. Of the 1,200 friend relationships investigated for the remaining 40 participants (30 friends per participant),  APM automatically labeled 403 as potentially abusive. AbuSniff predicted that 359 of these will be approved by the participants, i.e., 41 unfollow, 30 restrict, 137 unfriend and 151 sandbox. All the unfollow and 29 of the 30 restrict suggestions were accepted by the participants. 119 of the suggested sandbox relationships and 92 of the suggested unfriend relationships were accepted.  Thus, overall, the 40 participants accepted 78\% of AbuSniff's suggestions.

\subsection{Impact of AbuSniff}
\label{sec:evaluation:impact}

\begin{figure}
\centering
\subfigure[]
{\label{fig:pretest_vs_posttest}{\includegraphics[width=0.47\textwidth]{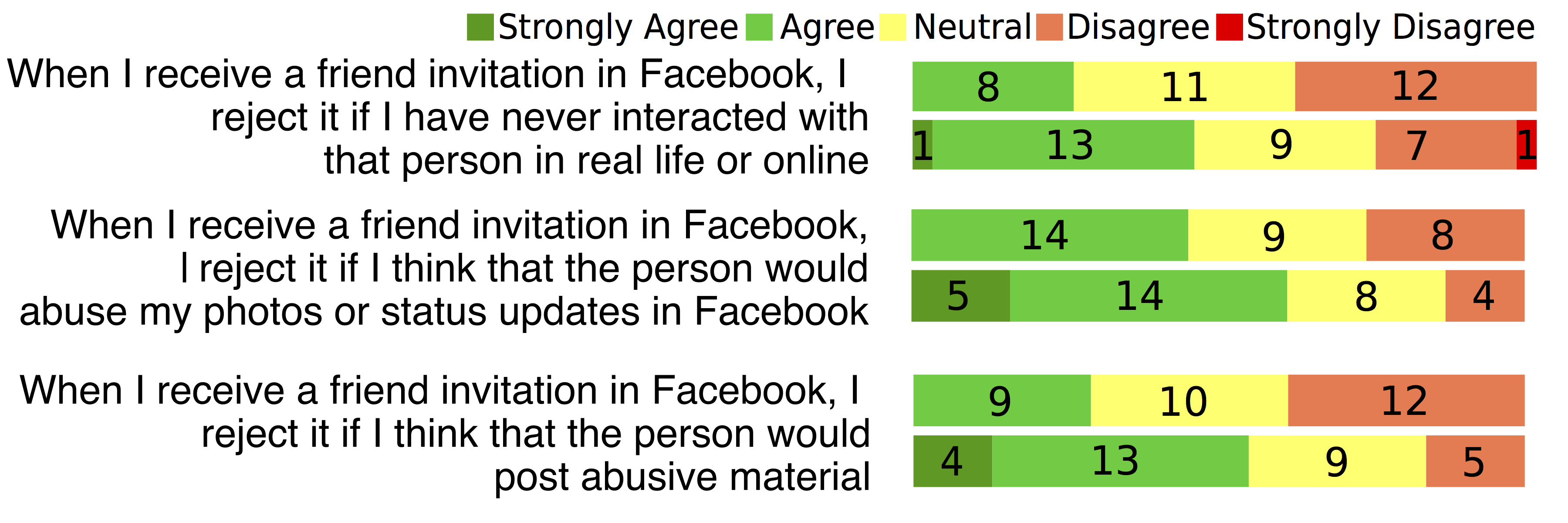}}}
\subfigure[]
{\label{fig:intervention}{\includegraphics[width=0.47\textwidth]{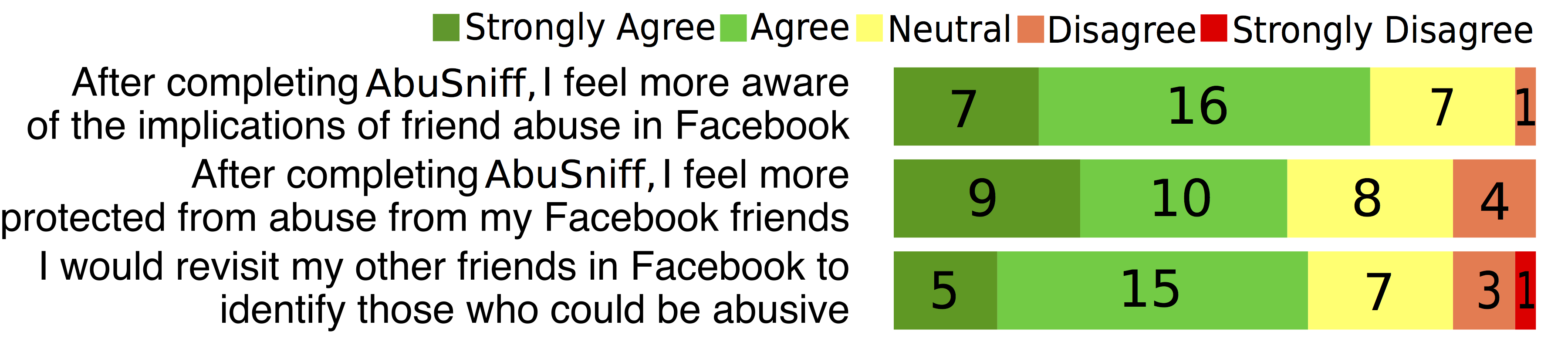}}}
\caption{
(a) AbuSniff impact on (I1), (I2) and (I3). For each question, top bar shows pre-test and bottom bar shows post-test results.
In the post-test, significantly more participants tend to strongly agree or agree that they would reject new friend invitations based on lack of interaction or perceived timeline or news feed abuse, when compared to the pre-test.
(b) Post-test results for (I4), (I5) and (I6).}
\vspace{-15px}
\end{figure}

In the last 2 user studies we have evaluated the impact of AbuSniff on (1) the willingness of participants to ignore new friend invitations based on their perception of the prospective friend being a stranger or an abuser, on (2) participant awareness of and perception of safety from friend abuse, and (3) their willingness to screen other friends.

For this, we have designed a pre-study survey that consists of 3 Likert items: (I1) ``When I receive a friend invitation in Facebook, I reject it if I have never interacted with that person in real life or online'', (I2) ``When I receive a friend invitation in Facebook, I reject it if I think that the person would abuse my photos or status updates in Facebook, and (I3) `` When I receive a friend invitation in Facebook, I reject it if I think that the person would post abusive material (offensive, misleading, false or potentially malicious).'' We performed a pre-test only study with 31 participants, where we have delivered (only) this survey.

Further, we have designed a post-study survey that consists of the above 3 items, plus the following 3 Likert items: (I4) ``After completing AbuSniff, I feel more aware of the implications of friend abuse in Facebook'', (I5) ``After completing AbuSniff, I feel more protected from abuse from Facebook friends'', and (I6) ``I will go to my friend list and evaluate my other friends to defend against those I feel could be abusive''. In a post-test study with a different set of 31 participants, we asked them to first run the questionnaire based AbuSniff version, then answer the post-study survey.

Figure~\ref{fig:pretest_vs_posttest} compares the user responses in the pre-test (top) and post-test (bottom) for each of the first 3 Likert items. In the pre-test, the user responses are balanced between agree, neutral and disagree, and there are no strong agree and strong disagree responses. In contrast, after running AbuSniff, significantly more participants either strongly agree or agree on all 3 items.

Figure~\ref{fig:intervention} shows the participant responses to only the 3 new post-test Likert items. 23 out of 31 participants strongly agree or agree that after running AbuSniff they feel more aware of the implications of friend abuse; only 1 disagreed. 19 participants strongly agree or agree that after running AbuSniff they feel more protected from friend abuse; 4 participants disagree. 20 participants strongly agree or agree that they would revisit their other friends after running AbuSniff, and only 3 disagree, 1 strongly disagrees.

\section{Discussion and Limitations}
\label{sec:discussion}

AbuSniff differs from prior work on cyber abuse and victimization, e.g.,~\cite{WCZBFEBSS16,LLL13} in that we (1) focus on specific types of abuse perpetrated through Facebook, i.e., timeline and news-feed abuse, and stranger friends, (2) investigate abuse perception from individual friends and not general exposure, (3) seek to automatically detect abuse perception and (4) provide a first line of suitable defenses against abuse for Facebook users who are unlikely to know and trust all their friends. We performed studies with 263 participants from 25 countries and 6 continents. We acknowledge a common crowdsourcing worker background between participants.

AbuSniff reduces the {\it attack surface} of its users, by reducing the number of, or isolating friends predicted to be perceived as potential attack vectors. AbuSniff can reduce the audience that needs to be considered by audience selector solutions, e.g.,~\cite{RLG16}, and can be used in conjunction with tools that monitor social networking events~\cite{DJHLP12,Perspective}.

We expect AbuSniff to have more impact for users who have significantly more than 150 friends, the maximum number of meaningful friend relationships that humans can manage~\cite{D92}. We note that false positives, while being a nuisance, can be fixed by reinstating removed or restricted friends. However, false negatives (keeping abusive and stranger friends) can harm the user and even influence the outcome of elections~\cite{BBCAbuse,BBCMay}.

\noindent
{\bf Online relationships and loose tie friends}.
Social networks like Facebook encourage online relationships (people never met in real life) and loose ties (users keeping up to date with the posts of others, without bi-directional communication). AbuSniff defines and detects ``strangers'' as friends with whom the user has no online and real world communications. Thus, since ``keeping up to date'' is considered interaction, AbuSniff does not detect and suggest removing strictly online relationships or loose tie friends.

\noindent
{\bf Prediction accuracy}.
The APM features extracted from mutual Facebook activities are less effective
in predicting the user responses to Q3-Q5. This is not surprising, as we have
trained APM on relationship closeness features. The choice of features was
needed to respect Facebook's terms of service. Access to more information,
e.g., stories on which friends posted replies and the friend replies, and abuse
detection APIs~\cite{Perspective} can improve APM's prediction performance. We
observe that AbuSniff had an F-Measure of 97.3\% when predicting the
``ignore'' action.

\noindent
{\bf Keeping friends perceived to be abusive}.
In the first study, for 11 of the 68 unfriended friend cases, the participants believed that our warning was correct, but still wanted to keep those friends. One reason may be that the participant had reasons to make him or her abusive toward that friend. We leave this investigation for future work, but note that AbuSniff may protect the friends if they installed AbuSniff.

\noindent
{\bf Friend evaluation limitations}.
We chose to evaluate 20 to 30 friends per participant. A larger number may increase participant fatigue or boredom when answering the questionnaire, thus reduce the quality of the data. More studies are needed to find the optimal number of evaluated friends per participant, and whether it should be a function of the participant background, e.g., friend count, age, gender.

\section{Related Work}

The features provided by online services are known to influence abuse and generate negative socio-psychological effects~\cite{SRHF17}. Social networks in particular enable a diverse set of abusive behaviors, that include the adversarial collection and abuse of private information~\cite{BBCAbuse,sextorsion,YS14,nosko2010all,GHWLCZ10}, cyberbullying~\cite{WCZBFEBSS16,LLL13,EN11,QBC12}, and the viral distribution of fake news, misinformation, propaganda and malware~\cite{BBCAbuse,BBCSenate,sindelar2014kremlin,aro2016cyberspace,weimann2010terror,al2010taking}.

\cite{CSYK15} detect the fake accounts behind friend spam, by extending the Kernighan-Lin heuristic to partition the social graph into two regions, that minimize the aggregate acceptance rate of friend requests from one region to the other. \cite{WSHY15} utilized posting relations between users and messages to combine social spammer and spam message detection. \cite{QH10} maintain information about friendly and suspicious devices that the user encounters in time, to decide if the user is the target of a friend spam attack. AbuSniff focuses on the user perception of strangers friends, their automatic detection and defenses.

\cite{WCZBFEBSS16} have used the Partner Cyber Abuse Questionnaire and found a prevalence of 40\% of victimization by cyber abuse among college students in dating relationships, with no differences in victimization of men and women. \cite{LLL13} developed the Social Networking-Peer Experiences Questionnaire (SN-PEQ) and used it to study cyber victimization in adolescents and young adults. They found that negative social networking experiences were associated with symptoms of social anxiety and depression. \cite{EN11} developed an 8 item questionnaire to explore the impact of involvement with Facebook on relationship satisfaction and found that Facebook intrusion was linked to relationship dissatisfaction via experiences of cognitive jealousy and surveillance behaviors. AbuSniff can help detect and protect against such behaviors.

\cite{QBC12} found that the reasons for ending friend relations are similar in the real and online worlds, and conjectured that tools can be built to monitor online relations. To detect cyberbullying, \cite{DJHLP12} used datasets of manually annotated comments, NLP features, supervised learning, and reasoning technique, then proposed several intervention designs. \cite{AV16} developed and conducted participatory design sessions with teenage participants to design, improve, and evaluate prototypes that address cyberbullying scenarios. \cite{NSLOOHS17} introduced an automated Twitter assistant that identifies text and visual bias, aggregates and presents evidence of bias to users, and enable activists to inform the public of bias, through bots.

\cite{KCM11} found that in Twitter, Korean users tended to unfollow people who
posted many tweets per time unit, about uninteresting topics, including details
of their lives. \cite{XHKC13} found that unfollow tends to be reciprocal. This
relatively harmless tit-for-tat behavior may explain the willingness of
participants in our studies to unfollow abusive friends. \cite{KML12} found
that users who receive acknowledgments from others are less likely to unfollow
them. Future work may compare the willingness of a user to unfollow Facebook
friends who posted general abuse vs. abuse personally targeted to the user.

\section{Conclusions}
\label{sec:conclusions}

We have introduced and studied AbuSniff, the first friend abuse detection and defense system for Facebook. We have developed a compact ``stranger and abuse'' detection questionnaire. We have introduced and studied rules to convert questionnaire answers to defensive actions. We have shown that supervised learning algorithms can use social networking based features to predict questionnaire answers and defense choices. AbuSniff increased participant willingness to reject invitations from perceived strangers and abusers, as well as awareness of friend abuse implications and perceived protection from friend abuse.

\section{Acknowledgments} 
\label{sec:acknowledgments}

We thank Mozhgan Azimpourkivi and Debra Davis for early discussions. This research was gracefully supported by NSF grants SES-1450619, CNS-1527153, and CNS-1526494 and by the Florida Center for Cybersecurity.

\bibliographystyle{aaai}
\fontsize{9.5pt}{10.5pt}
\selectfont
\bibliography{talukder}

\begin{thebibliography}{}

\bibitem[\protect\citeauthoryear{Al-Shishani}{2010}]{al2010taking}
Al-Shishani, M.~B.
\newblock 2010.
\newblock Taking al-qaeda's jihad to facebook.
\newblock {\em The Jamestown Foundation: Terrorism Monitor} 8(5):3.

\bibitem[\protect\citeauthoryear{Angwin and Grassegger}{2017}]{CNBCabuse}
Angwin, J., and Grassegger, H.
\newblock 2017.
\newblock Facebook’s secret censorship rules protect white men from hate
  speech but not black children.
\newblock [CNBC Tech] tinyurl.com/y7ncjgqx.

\bibitem[\protect\citeauthoryear{Aro}{2016}]{aro2016cyberspace}
Aro, J.
\newblock 2016.
\newblock The cyberspace war: propaganda and trolling as warfare tools.
\newblock {\em European View} 15(1).

\bibitem[\protect\citeauthoryear{Ashktorab and Vitak}{2016}]{AV16}
Ashktorab, Z., and Vitak, J.
\newblock 2016.
\newblock Designing cyberbullying mitigation and prevention solutions through
  participatory design with teenagers.
\newblock In {\em Proceedings of CHI}.

\bibitem[\protect\citeauthoryear{BBC}{2017a}]{BBCAbuse}
BBC.
\newblock 2017a.
\newblock Russia-linked posts \'reached 126m facebook users in us\'.
\newblock [BBC News] tinyurl.com/y93ylwdw.

\bibitem[\protect\citeauthoryear{BBC}{2017b}]{BBCMay}
BBC.
\newblock 2017b.
\newblock {Theresa May accuses Vladimir Putin of election meddling}.
\newblock [BBC Politics] tinyurl.com/y8d2pwmy.

\bibitem[\protect\citeauthoryear{Cao \bgroup et al\mbox.\egroup
  }{2015}]{CSYK15}
Cao, Q.; Sirivianos, M.; Yang, X.; ; and Munagala, K.
\newblock 2015.
\newblock {Combating Friend Spam Using Social Rejections}.
\newblock In {\em Proceedings of the IEEE ICDCS}.

\bibitem[\protect\citeauthoryear{Cheng \bgroup et al\mbox.\egroup
  }{2017}]{CBDL17}
Cheng, J.; Bernstein, M.~S.; Danescu{-}Niculescu{-}Mizil, C.; and Leskovec, J.
\newblock 2017.
\newblock Anyone can become a troll: Causes of trolling behavior in online
  discussions.
\newblock In {\em {ACM CSCW}}.

\bibitem[\protect\citeauthoryear{Dinakar \bgroup et al\mbox.\egroup
  }{2012}]{DJHLP12}
Dinakar, K.; Jones, B.; Havasi, C.; Lieberman, H.; and Picard, R.
\newblock 2012.
\newblock Common sense reasoning for detection, prevention, and mitigation of
  cyberbullying.
\newblock {\em ACM TiiS}.

\bibitem[\protect\citeauthoryear{Dunbar}{1992}]{D92}
Dunbar, R.~I.
\newblock 1992.
\newblock Neocortex size as a constraint on group size in primates.
\newblock {\em Journal of human evolution} 22(6):469--493.

\bibitem[\protect\citeauthoryear{Elphinston and Noller}{2011}]{EN11}
Elphinston, R.~A., and Noller, P.
\newblock 2011.
\newblock Time to face it! facebook intrusion and the implications for romantic
  jealousy and relationship satisfaction.
\newblock {\em Cyberpsychology, Behavior, and Social Networking}
  14(11):631--635.

\bibitem[\protect\citeauthoryear{Gao \bgroup et al\mbox.\egroup
  }{2010}]{GHWLCZ10}
Gao, H.; Hu, J.; Wilson, C.; Li, Z.; Chen, Y.; and Zhao, B.~Y.
\newblock 2010.
\newblock Detecting and characterizing social spam campaigns.
\newblock In {\em {Proceedings of the ACM IMC}}.

\bibitem[\protect\citeauthoryear{Heath}{2017}]{botsandclones}
Heath, A.
\newblock 2017.
\newblock Facebook quietly updated two key numbers about its user base.
\newblock [Business Insider] tinyurl.com/y76s8rvs.

\bibitem[\protect\citeauthoryear{Job}{2017}]{JobBoy}
2017.
\newblock {JobBoy}.
\newblock http://www.jobboy.com/.

\bibitem[\protect\citeauthoryear{Kontaxis \bgroup et al\mbox.\egroup
  }{2011}]{kontaxis2011detecting}
Kontaxis, G.; Polakis, I.; Ioannidis, S.; and Markatos, E.~P.
\newblock 2011.
\newblock Detecting social network profile cloning.
\newblock In {\em {PercomW}}.

\bibitem[\protect\citeauthoryear{Kwak, Chun, and Moon}{2011}]{KCM11}
Kwak, H.; Chun, H.; and Moon, S.
\newblock 2011.
\newblock {Fragile Online Relationship: A First Look at Unfollow Dynamics in
  Twitter}.
\newblock In {\em Proceedings of ACM CHI},  1091--1100.

\bibitem[\protect\citeauthoryear{Kwak, Moon, and Lee}{2012}]{KML12}
Kwak, H.; Moon, S.~B.; and Lee, W.
\newblock 2012.
\newblock {More of a receiver than a giver: why do people unfollow in Twitter?}
\newblock In {\em Proceedings of the AAAI ICWSM}.

\bibitem[\protect\citeauthoryear{Kwan and Skoric}{2013}]{kwan2013facebook}
Kwan, G. C.~E., and Skoric, M.~M.
\newblock 2013.
\newblock Facebook bullying: An extension of battles in school.
\newblock {\em Computers in Human Behavior} 29(1):16--25.

\bibitem[\protect\citeauthoryear{Landoll, La~Greca, and Lai}{2013}]{LLL13}
Landoll, R.~R.; La~Greca, A.~M.; and Lai, B.~S.
\newblock 2013.
\newblock Aversive peer experiences on social networking sites: Development of
  the social networking-peer experiences questionnaire (sn-peq).
\newblock {\em Journal of Research on Adolescence} 23(4):695--705.

\bibitem[\protect\citeauthoryear{Lapowsky}{2018}]{CAWired}
Lapowsky, I.
\newblock 2018.
\newblock {Cambridge Analytica Execs Caught Discussing Extorsion and Fake
  News}.
\newblock [Wired] https://tinyurl.com/yaagbe9h.

\bibitem[\protect\citeauthoryear{Lee}{2017}]{BBCSenate}
Lee, D.
\newblock 2017.
\newblock {Facebook, Twitter and Google berated by senators on Russia}.
\newblock [BBC Technology] tinyurl.com/ybmd55js.

\bibitem[\protect\citeauthoryear{Madejski, Johnson, and Bellovin}{2012}]{MJB12}
Madejski, M.; Johnson, M.; and Bellovin, S.~M.
\newblock 2012.
\newblock A study of privacy settings errors in an online social network.
\newblock In {\em {Proceedings of PERCOM Workshops}}.

\bibitem[\protect\citeauthoryear{Narwal \bgroup et al\mbox.\egroup
  }{2017}]{NSLOOHS17}
Narwal, V.; Salih, M.~H.; Lopez, J.~A.; Ortega, A.; O'Donovan, J.;
  H\"{o}llerer, T.; and Savage, S.
\newblock 2017.
\newblock Automated assistants to identify and prompt action on visual news
  bias.
\newblock In {\em ACM CHI}.

\bibitem[\protect\citeauthoryear{Nosko, Wood, and Molema}{2010}]{nosko2010all}
Nosko, A.; Wood, E.; and Molema, S.
\newblock 2010.
\newblock All about me: Disclosure in online social networking profiles: The
  case of facebook.
\newblock {\em Computers in Human Behavior} 26(3).

\bibitem[\protect\citeauthoryear{Ortutay and Jesdanun}{2018}]{CAABC}
Ortutay, B., and Jesdanun, A.
\newblock 2018.
\newblock {How Facebook likes could profile voters for manipulation}.
\newblock [ABC News] https://tinyurl.com/yaaf3lws.

\bibitem[\protect\citeauthoryear{Per}{2017}]{Perspective}
2017.
\newblock {What if technology could help improve conversations online?}
\newblock https://www.perspectiveapi.com/.

\bibitem[\protect\citeauthoryear{Quercia and Hailes}{2010}]{QH10}
Quercia, D., and Hailes, S.
\newblock 2010.
\newblock Sybil attacks against mobile users: friends and foes to the rescue.
\newblock In {\em IEEE INFOCOM}.

\bibitem[\protect\citeauthoryear{Quercia, Bodaghi, and Crowcroft}{2012}]{QBC12}
Quercia, D.; Bodaghi, M.; and Crowcroft, J.
\newblock 2012.
\newblock {Loosing "friends" on Facebook}.
\newblock In {\em ACM WebSci}.

\bibitem[\protect\citeauthoryear{Raber, Luca, and Graus}{2016}]{RLG16}
Raber, F.; Luca, A.~D.; and Graus, M.
\newblock 2016.
\newblock Privacy wedges: Area-based audience selection for social network
  posts.
\newblock In {\em Proceedings of SOUPS}.

\bibitem[\protect\citeauthoryear{Shahani}{2016}]{NPRabuse}
Shahani, A.
\newblock 2016.
\newblock {From Hate Speech To Fake News: The Content Crisis Facing Mark
  Zuckerberg}.
\newblock [NPR] tinyurl.com/ycxmke8r.

\bibitem[\protect\citeauthoryear{Sindelar}{2014}]{sindelar2014kremlin}
Sindelar, D.
\newblock 2014.
\newblock The kremlin's troll army.
\newblock {\em The Atlantic} 12.

\bibitem[\protect\citeauthoryear{Singh \bgroup et al\mbox.\egroup
  }{2017}]{SRHF17}
Singh, V.~K.; Radford, M.~L.; Huang, Q.; and Furrer, S.
\newblock 2017.
\newblock {"They basically like destroyed the school one day": On Newer App
  Features and Cyberbullying in Schools}.
\newblock In {\em {ACM CSCW}}.

\bibitem[\protect\citeauthoryear{Smith}{2008}]{MS08}
Smith, M.
\newblock 2008.
\newblock {Facebook Abuse: Is Blocking People Enough?}
\newblock tinyurl.com/yajt93gh.

\bibitem[\protect\citeauthoryear{Varol \bgroup et al\mbox.\egroup
  }{2017}]{VFDMF17}
Varol, O.; Ferrara, E.; Davis, C.~A.; Menczer, F.; and Flammini, A.
\newblock 2017.
\newblock Online human-bot interactions: Detection, estimation, and
  characterization.
\newblock In {\em Proceedings of the AAAI ICWSM}.

\bibitem[\protect\citeauthoryear{Vitak and Kim}{2014}]{VK14}
Vitak, J., and Kim, J.
\newblock 2014.
\newblock {You can't block people offline": examining how Facebook's
  affordances shape the disclosure process}.
\newblock In {\em Proceedings of CSCW}.

\bibitem[\protect\citeauthoryear{Wakefield}{2017}]{BBCUK}
Wakefield, J.
\newblock 2017.
\newblock {Facebook and Twitter could face 'online abuse levy'}.
\newblock [BBC Technology] tinyurl.com/ycob8538.

\bibitem[\protect\citeauthoryear{Weimann}{2010}]{weimann2010terror}
Weimann, G.
\newblock 2010.
\newblock Terror on facebook, twitter, and youtube.
\newblock {\em The Brown Journal of World Affairs} 16(2).

\bibitem[\protect\citeauthoryear{Wek}{2017}]{Weka}
2017.
\newblock {Weka}.
\newblock tinyurl.com/36z952.

\bibitem[\protect\citeauthoryear{Wolford-Cleveng \bgroup et al\mbox.\egroup
  }{2016}]{WCZBFEBSS16}
Wolford-Cleveng, C.; Zapor, H.; Brasfield, H.; Febres, J.; Elmquist, J.; Brem,
  M.; Shorey, R.~C.; and Stuart, G.~L.
\newblock 2016.
\newblock An examination of the partner cyber abuse questionnaire in a college
  student sample.
\newblock {\em Psychology of violence} 6(1):156.

\bibitem[\protect\citeauthoryear{Wu \bgroup et al\mbox.\egroup }{2015}]{WSHY15}
Wu, F.; Shu, J.; Huang, Y.; and Yuan, Z.
\newblock 2015.
\newblock Social spammer and spam message co-detection in microblogging with
  social context regularization.
\newblock In {\em Proceedings of the ACM CIKM}.

\bibitem[\protect\citeauthoryear{Xu \bgroup et al\mbox.\egroup }{2013}]{XHKC13}
Xu, B.; Huang, Y.; Kwak, H.; and Contractor, N.
\newblock 2013.
\newblock Structures of broken ties: Exploring unfollow behavior on twitter.
\newblock In {\em Proceedings of ACM CSCW},  871--876.

\bibitem[\protect\citeauthoryear{Yang and Srinivasan}{2014}]{YS14}
Yang, C., and Srinivasan, P.
\newblock 2014.
\newblock Translating surveys to surveillance on social media: methodological
  challenges {\&} solutions.
\newblock In {\em {Proceedings of WebSci}}.

\bibitem[\protect\citeauthoryear{Yates}{2017}]{sextorsion}
Yates, J.
\newblock 2017.
\newblock {From temptation to sextortion Inside the fake Facebook profile
  industry}.
\newblock [Radio Canada] tinyurl.com/ycf5md7t.

\end{thebibliography}

\end{document}